\documentclass[onecolumn,amsmath,amssymb]{qm2008_abs}
\usepackage{graphicx}
\usepackage{dcolumn}
\usepackage{bm}
\begin{document}

\title{{Neutral pion production with respect to reaction plane in
    Au+Au collisions at RHIC-PHENIX}}

\bigskip
\bigskip
\author{\large Yoki Aramaki for the PHENIX Collaboration}
\email{aramaki@cns.s.u-tokyo.ac.jp}
\affiliation{University of Tokyo, Tokyo, Japan} 
\bigskip
\bigskip

\begin{abstract}
\leftskip1.0cm
\rightskip1.0cm
We have measured azimuthal anisotropy of neutral pion in
$\sqrt{s_{NN}}$=200~GeV Au+Au collisions at Relativistic Heavy Ion
Collider (RHIC).  
In 2007, we have installed a new detector to measure reaction plane of
collisions more precisely.  
This new detector is called reaction plane detector (RxNP).  
We report the results for azimuthal anisotropy of neutral pion for
each 3 centrality steps.  
The determination of reaction plane achieved twice better resolution
than Year-2004 run.     
\end{abstract}

\maketitle

\section{Introduction}
It has been observed in central Au+Au collisions at Relativistic Heavy
Ion Collider (RHIC) that the yield of neutral pions at high transverse
momentum ($p_{T}$$>$5~GeV/$c$) is strongly suppressed compared to
the one expected from p+p collisions.  
This suppression is considered to be an energy loss of hard scattered
partons in the medium (jet quenching), that results in a decrease of
the yield at a given $p_{T}$\cite{jet_quenching}.   
The magnitude of the suppression depends on the path length of
scattering partons in the medium, and as shown in
FIG.~\ref{fig:scheme_rp}, impact parameter is experimentally associated
with the azimuthal angle of emitted particles from the reaction plane
in non-central collisions.
Path length can be calculated from Glauber model and impact parameter.
Studying the path length dependence of energy loss should provide
additional information on the energy loss mechanism in the medium.  
Some theoretical models suggest that LPM effect in quantum
chromodynamics (QCD) plays an important role in energy loss
mechanism\cite{lpm}.    
LPM effect in QCD is correlated with path length that partons pass
through the medium.  
We discuss the parton energy loss mechanism using the nuclear
modification factor ($R_{AA}$) of neutral pion with respect to
reaction plane.  
The nuclear modification factor can be expressed using centrality
(cent) which is associated to impact parameter for collisions and
$p_{T}$.     

\begin{eqnarray}
R_{AA}(p_{T}, cent) = \frac{\sigma^{inel}_{pp}}{\langle
  N_{coll}\rangle}\frac{d^{2}N^{AA}/dp_{T}d\eta}{d^{2}\sigma^{NN}/dp_{T}d\eta},   
\end{eqnarray}  
where $N^{AA}$ is the number of neutral pions in a given centrality.
$\sigma^{inel}_{pp}$ is a cross section for inelastic nucleon-nucleon
collisions.  
A value of $R_{AA}(p_{T}, cent)= 1$ implies that particle production
is scaled by the average number of binary nucleon-nucleon collisions,
$\langle N_{coll} \rangle$.  The modification factor for each
azimuthal angles is expressed by the following equation.  

\begin{eqnarray}
R_{AA}(p_{T}, cent, \Delta\phi) = \frac{N^{AA}(\Delta\phi)}{\int d\phi
  N^{AA}(\Delta\phi)}\times R_{AA}(p_{T}, cent),
\end{eqnarray}
where $N^{AA}$ can be expressed in terms of a Fourier expansion with
$\Delta\phi$.  

\begin{eqnarray}
N^{AA}(\Delta\phi) \propto 1+2\sum_{n=1}^{\infty}v_{n}\cos(n\Delta\phi), 
\label{eq:dndphi}
\end{eqnarray}
where $v_{2}$ is the strength of azimuthal anisotropy.

A new reaction plane detector, called as RxNP, was installed in the
PHENIX experiment in the Year-7 run, and expected to improve the
reaction plane resolution.  
Additionally we obtained 3.5 times higher statistics than RHIC Year-4.   
The integrated luminosity in Year-7 run is 813~$\mu b$.  
Much precise measurement of the hadron suppression with respect to
path length is expected using the detector.  
Azimuthal anisotropy of neutral pions ($\pi^{0}\rightarrow 2\gamma$)
is investigated in this report, since measurement of neutral pions up
to $\sim$20~GeV/$c$ is possible with the PHENIX electromagnetic
calorimeters (EMCal).

\section{Experimental Setup}
Reaction plane of collisions has determined using with the Beam-Beam
Counters (BBC) before installation of the new detector (RxNP).  
RxNP was designed to have a wider rapidity coverage (1.0$<$ $|\eta|$
$<$1.5 $\&$ 1.5$<$ $|\eta|$ $<$ 2.8) compared with BBC (3$<$ $|\eta|$
$<$4). 
Better resolution in reaction plane determination is achieved with this
wider coverage.  
PHENIX has two types of calorimeters.  
PbSc,  consists of the lead and scintillator plates and wavelength
shifter fiber readout.   
Another type is lead-glass calorimeter (PbGl).  
The PbSc has a nominal energy and position resolution of
$8.1/\sqrt{E(GeV)}\bigoplus 2.1~\%$ and $5.7/\sqrt{E(GeV)} \bigoplus
21.55~mm$, respectively. 
The PbGl has a nominal energy resolution of $5.9/\sqrt{E(GeV)}
\bigoplus0.8~ \%$ and $8.4/\sqrt{E(GeV)}\bigoplus 0.2~mm$,
respectively.      
 
\section{Analysis and Results}
\begin{figure}[htbp]
  \centering
   \includegraphics[width=7cm]{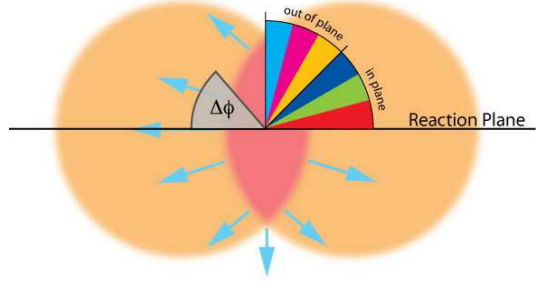}
    \caption{Schematic of the produced medium.  In this analysis,
     azimuthal angles from reaction plane are divided into six azimuthal angles.}
    \label{fig:scheme_rp}
\end{figure}

\begin{figure}[htbp]
  \begin{minipage}{.25\linewidth}
    \includegraphics[width=\linewidth]{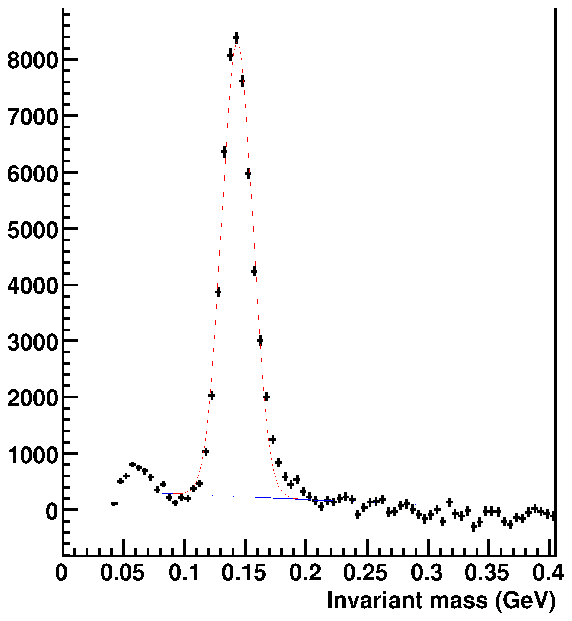} 
    \caption{Subtracted 2$\gamma$ invariant mass distributions.}
    \label{fig:invmass}
  \end{minipage}
  \vspace{3cm}
  \begin{minipage}{.3\linewidth}
    \includegraphics[width=\linewidth]{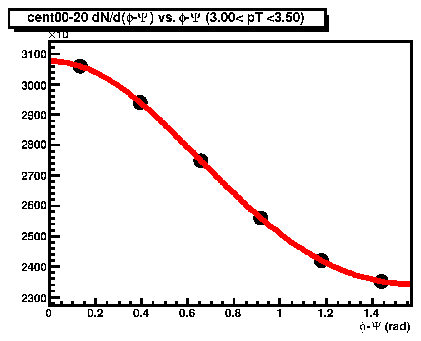}
     \caption{Counted $\pi^{0}$'s distributions for each azimuthal
     angles in 3$<$$p_{T}$$<$3.5 GeV/$c$ (Centrality 0-20$\%$).} 
    \label{fig:dndphi}
    \end{minipage}
  \begin{minipage}{.4\linewidth}
    \includegraphics[width=\linewidth]{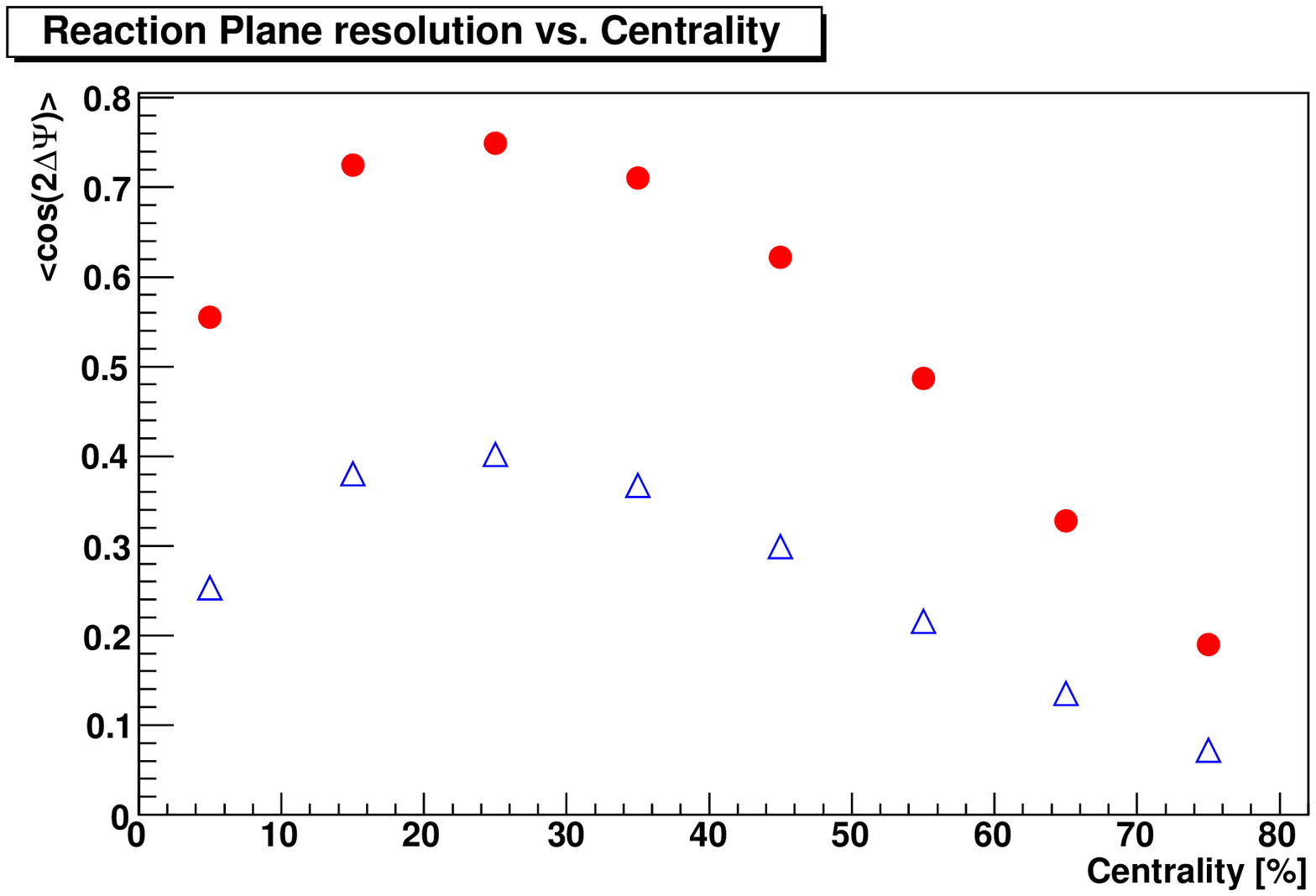}
     \caption{Comparison of reaction plane resolution for RxNP and BBC.
     Circular points and triangular points are shown reaction plane
     resolution for RxNP and BBC, respectively.} 
       \label{fig:RPreso}
    \end{minipage}
\end{figure}

\noindent
We selected events with a vertex position within $\pm$ 30~cm on
z-axis.
For each selected event, made was determination of reaction plane.
Reaction plane is experimentally determined by the direction which the
greatest number of particles are emitted.  
As shown in FIG.~\ref{fig:invmass}, neutral pions are identified by
2~$\gamma$ invariant mass calculated in same event.
The combinatorial background is evaluated with event mixing technique
which 2 $\gamma$ candidates in different events are calculated.  
Neutral pions are counted in a given mass window (Typically 2~$\sigma$
from peak position of 2~$\gamma$ invariant mass distribution.).  
As shown in FIG.~\ref{fig:scheme_rp}, the number of counted neutral
pions are divided into six $\Delta\phi$ bins in the interval from 0 to
$\pi/2$. 

$\Delta\phi$ means the angle which are subtracted the direction to the
reaction plane from the angular direction to the emitted particles.  
This reaction plane can be determined by BBC or RxNP.  
As shown in FIG.~\ref{fig:dndphi}, we fit the following eq.(\ref{eq:dndphi_appro}).

\begin{eqnarray}
f(\Delta\phi) =
N_{0}(1+2v^{raw}_{2}\cos(2\Delta\phi)+2v^{raw}_{4}\cos(4\Delta\phi)),
\label{eq:dndphi_appro}
\end{eqnarray}
where $v^{raw}_{2}$ is observed raw azimuthal anisotropy.  
The true azimuthal anisotropy $v^{corr}_{2}$ needs to be corrected by
the reaction plane resolution\cite{rp_reso}.
\begin{eqnarray}
v^{raw}_{2} = v^{corr}_{2}\langle\cos(2\Delta\Psi)\rangle,
\end{eqnarray}
where $\Delta\Psi$ is a difference between the true azimuthal angle
from reaction plane and the observed angle from reaction plane.  
As shown in FIG.~\ref{fig:RPreso}, the new detector (RxNP) could
improve this $\langle\cos(2\Delta\phi)\rangle$ value.

\begin{figure}[htbp]
  \centering
   \includegraphics[width=15cm]{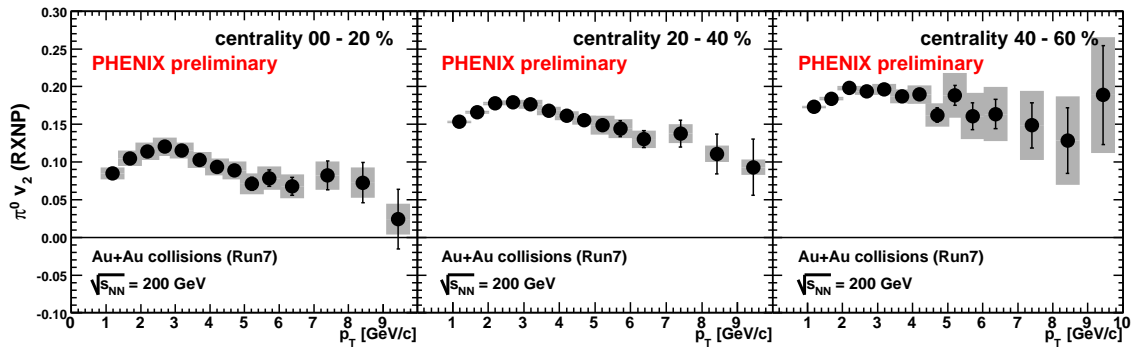}
    \caption{Run~7 $\pi^{0} v_2$ with RxNP result for 20$\%$
    centrality steps.  The statistics is $1/3$ of the totally available.}
    \label{fig:RXNPv2_cent00-60}
\end{figure}
Systematic uncertainty sources for this measurement are $\pi^{0}$'s
counting uncertainty and uncertainty for $v_{2}$ difference between
BBC and RxNP.  
Both of BBC and RxNP can measure reaction plane of collisions.  
Thus $\pi^{0} v_{2}$ difference between these detectors is taken into
account as a systematic uncertainty.  
The evaluation for $v_{2}$ difference has been done by measurement of
charged pion $v_{2}$ with BBC and RxNP.  
For centrality 0-20$\%$ step, systematic uncertainty of the $\pi^{0}$
counting (1$<$$p_{T}$$<$3, 3$<$$p_{T}$$<$5, 5$<$$p_{T}$$<$10~GeV/$c$)
is 9$\%$, 7$\%$ and 12$\%$, respectively. 
Systematic uncertainty of reaction plane determination (1$<$$p_{T}$$<$3,
3$<$$p_{T}$$<$5, 5$<$$p_{T}$$<$10~GeV/$c$) is 4$\%$, 9$\%$ and 29.6$\%$,
respectively.  Hence total systematic uncertainty (1$<$$p_{T}$$<$3,
3$<$$p_{T}$$<$5, 5$<$$p_{T}$$<$10~GeV/$c$) is 10$\%$, 11$\%$ and 32$\%$,
respectively.

As shown in FIG.~\ref{fig:RXNPv2_cent00-60}, azimuthal anisotropy of
neutral pion still remains a finite value even in high $p_{T}$.  
Now we have measured $\pi^{0} v_{2}$ up to $p_{T}\sim$10~GeV/$c$ using
$1/3$ of totally available.  
Thus measurement of $\pi^{0} v_{2}$ can be done up to higher $p_{T}$
region.  
The path length dependence of $\pi^{0} R_{AA}$ needs to measure in
smaller centrality steps.  
More precisely study of path length dependence of $\pi^{0} R_{AA}$ can
be done by better reaction plane resolution and higher statistics than the
Year-4 run.
 
\section{Summary}
Measurement of azimuthal anisotropy of neutral pion has been done by
$1/3$ of total data using new detector (RxNP).  
This detector has achieved twice better reaction plane resolution
than with BBC.   
The values of $\pi^{0} v_{2}$ with BBC and RxNP still remains finite
up to $p_{T}\sim$10~GeV/$c$.

For studying the path length dependence of $\pi^{0}$ suppression, we
need to divide centrality into smaller bins.  
The modification factor for each path lengths can verify validity of
LPM effect in QCD for parton energy loss models. 
Thus, we can understand the parton energy loss mechanism more
quantitatively.


\medskip
\medskip


\noindent
\begin{thebibliography}{50}
\medskip
\bibitem{jet_quenching}S~S.Adler {\it et al.}, Phys. Rev. Lett {\bf
  96} (2006) 202301
\bibitem{lpm}X.~Wang, M.~Glulassy and M.~pl\"{u}mer, 
Phys. Rev. D {\bf 51} (1995) 3436.
\bibitem{rp_reso}J.~Barrette {\it et al}, Phys. Rev. C {\bf 55} (1997) 1420  

\end{thebibliography}
\end{document}